\documentstyle[12pt]{article}

\begin{document}
\begin{center}
{\Large \bf Relations between the scales of length, time and mass }
\bigskip

{\large D.L.~Khokhlov}
\smallskip

{\it Sumy State University, R.-Korsakov St. 2, \\
Sumy 40007, Ukraine\\
E-mail: khokhlov@cafe.sumy.ua}
\end{center}

\begin{abstract}
It is considered the model of the homogeneous and isotropic universe.
The scale of length is defined via the laboratory scale of time
by the motion of photon. This leads to the appearance of the
inertial forces. The properties of the space and time
are defined both by these inertial forces and by the matter.
Within the framework of classical physics, the scales of length,
time and mass are related by the special relativity constant $c$
and by the Newton gravity constant $G$.
Within the framework of quantum mechanics,
the scales of length, time and energy are related by
the special relativity constant $c$
and by the quantum mechanics constant $\hbar$.
The model meets constraints from the current age of the universe,
from the high-redshift supernovae data,
and from primordial nucleosynthesis.
The model predicts the fractal galaxy distribution with
a power index of 2.
\end{abstract}

\section{Introduction}

According to the general relativity the properties of the space
and time are defined by the matter. However in order to
specify the scales of length and time it is necessary to
introduce inertial reference frames, the rest frame and the
frame moving with the velocity $v$ relative to the rest frame.
While defining the scale of time $t$ in the rest frame,
one can introduce the scale of length
by means of the moving frame $R=vt$.
The transition from the rest frame to the moving frame specifies
an inertial acceleration $F_{in}=v/t=v^2/R$.
Thus the introduction of the space and time
by means of the inertial frames is accompanied by
the introduction of the non-inertial frames. This means
that a priori granted cosmological inertial forces
must exist in the universe.
Hence the properties of the space and time are defined both
by the matter and by the inertial forces.

\section{Theory}

\subsection{The space and time of the universe}

Consider the homogeneous and isotropic universe
in the laboratory system of reference.
Introduce the laboratory time $t$.
At the fixed moment of the laboratory time $t=const$,
the motion of photon relative to the rest frame
specifies the scale of length
\begin{equation}
a=ct.
\label{eq:at}
\end{equation}
Along this scale of length, the motion of photon specifies
the field of velocities
\begin{equation}
v=\frac{R}{a}c
\label{eq:vc}
\end{equation}
and the corresponding field of accelerations
\begin{equation}
F_{in}=\frac{v}{t}=\frac{v^2}{R}=\frac{c^2}{a^2}R.
\label{eq:Fin}
\end{equation}
The acceleration $F_{in}$ is due to the inertial force arising
under the transition from the rest frame to the moving frame.
Thus determination of the scale of length by means of photon
moving relative to the rest frame leads to the
appearance of the inertial acceleration.

Take the volume of the radius $a$. The matter homogeneously
distributed in this volume specifies the field of accelerations
due to gravity
\begin{equation}
F_{gr}=-G \rho R.
\label{eq:acc}
\end{equation}
The acceleration due to gravity $F_{gr}$ and the inertial
acceleration $F_{in}$ act in the opposite directions.
If the density of the matter is equal to
\begin{equation}
\rho =\frac{c^2}{Ga^2}
\label{eq:rhc}
\end{equation}
the acceleration due to gravity $F_{gr}$ balances the inertial
acceleration $F_{in}$, and the resulting acceleration
is equal to zero
\begin{equation}
F=F_{in} + F_{gr}=0.
\label{eq:acr}
\end{equation}
Thus, at $t=const$, the scale of length $a$ is the element of
the euclidean space,
with the gravitational forces due to the matter
of the universe balancing
the inertial forces due to the motion of photon.

It should be noted that the acceleration $F=\Lambda c^2 R$
proposed by Einstein~\cite{Zeld} coincides with the inertial
acceleration given by~(\ref{eq:Fin}) if to adopt $\Lambda=a^{-2}$.
Thus we arrive at the model of the universe in the spirit
of Einstein. Unlike the conventional point of view when
$\Lambda$-term is associated with the vacuum, here $\Lambda$-term
is associated with the cosmological inertial forces.
Unlike the Einstein model describing the static universe,
here the universe evolves following the linear law~(\ref{eq:at}).

\subsection{Evolution of the universe
within the framework of classical physics}

Take the universe as a particle with the radius $a$ and the mass $m$.
The values $a$ and $m$ are the scales of length and mass at the fixed
moment of the laboratory time $t=const$. Let the scale of length $a$
be the element of the euclidean space. That is the acceleration
due to gravity~(\ref{eq:acc}) balances the inertial
acceleration~(\ref{eq:Fin}). Then the density of the matter
in the universe is given by eq.~(\ref{eq:rhc}) that
corresponds to the closed world with zero total mass.
That is the mass of the matter of the universe is equal to the gravity
of the universe
\begin{equation}
mc^2=\frac{Gm^2}{a}.
\label{eq:c2}
\end{equation}

Evolution of the universe is defined by the motion of photon.
The law of evolution for the scale of length is given by
eq.~(\ref{eq:at}).
Since the scale of mass is bounded with the scale of length by
the relation~(\ref{eq:c2}), from the law of evolution for the scale of
length~(\ref{eq:at}) it follows the law of evolution for the scale of mass
\begin{equation}
m=\frac{c^2}{G}a=\frac{c^3}{G}t.
\label{eq:mt}
\end{equation}
Thus, within the framework of classical physics,
the scales of length and mass are the linear functions of time
which are defined by the constant of special relativity $c$ and
by the constant of the Newton gravity $G$.
The law of evolution for the scale of length~(\ref{eq:at}) and
the law of evolution for the scale of mass~(\ref{eq:mt}) give
the law of evolution for the density of the matter
\begin{equation}
\rho={{3m}\over{4\pi a^3}}={{3c^2}\over{4\pi G a^2}}={3\over{4\pi G t^2}}.
\label{eq:q}
\end{equation}

\subsection{Evolution of the universe
within the framework of quantum mechanics}

Consider the evolution of the universe
within the framework of quantum mechanics.
At $t=const$, compare the universe of the mass $m$ and the radius $a$
with a wave of the frequency
\begin{equation}
\omega\equiv\frac{E}{\hbar}=\frac{mc^2}{\hbar}
\label{eq:om}
\end{equation}
and of the wave number
\begin{equation}
{\rm k}\equiv\frac{\omega}{c}=\frac{1}{a}.
\label{eq:k}
\end{equation}
The frequency and the wave number are the scales of energy and length
respectively.

Due to the Heisenberg uncertainty principle
the scale of energy (frequency) and the lifetime of the unvierse
are related as
\begin{equation}
Et=\hbar,
\label{eq:Et}
\end{equation}
the scale of momentum projection $p_x$ and the scale of length $a_x$
are related as
\begin{equation}
p_x a_x=\hbar.
\label{eq:px}
\end{equation}
Thus, within the framework of quantum mechanics,
the evolution of the universe is governed by
the Heisenberg uncertainty principle. The evolution laws
for the scales of energy (frequency) and length are defined by
the constant of special relativity $c$ and by the constant
of quantum mechanics $\hbar$.
The evolution law for scale of length is linear like
in the classical physics.
Unlike the linear law of evolution for the scale of
mass in the classical physics, in the quantum mechanics
the law of evolution for the scale of energy (frequency)
is inversely linear. Hence one can
put the scale of energy (frequency) into correspondence to
the scale of mass only in a unique moment of time. Such a moment
when the scale of energy (mass) of the classical physics and
the scale of energy (frequency) of the quantum mechanics are
the same is the Planck time. Thus eq.~(\ref{eq:om}) holds true
at the Planck time.

Since the change of length is defined only in the direction $x$,
the evolution of the density of the matter is given by
\begin{equation}
\rho\propto E a{_x}^{-1}\propto a_{x}^{-2}\propto t^{-2}.
\label{eq:rhx}
\end{equation}
Thus the evolution of the density of the matter
is defined by the same law
both within the framework of classical physics (\ref{eq:q})
and within the framework of quantum mechanics (\ref{eq:rhx}).

At $t=const$, the density of the relativistic matter is expressed via
the scale of length and correspondingly via the temperature as
\begin{equation}
\rho\equiv a^{-4}\equiv T^{4}.
\label{eq:rha}
\end{equation}
Due to the evolution of the density of the matter (\ref{eq:rhx}),
the temperature defined from the density of the relativistic matter
changes with time as
\begin{equation}
T\equiv \rho^{1/4} \propto a^{-1/2}\propto t^{-1/2}.
\label{eq:tem}
\end{equation}

\subsection{The radial, angular diameter and luminosity distances}

The law for the velocity along the scale of length~(\ref{eq:vc})
leads to the invariance of the relations between the distances
under time transformation $t \rightarrow t+const$, $r_1/r_2=const$.
Since the scales of length and mass are bounded by the linear
relation~(\ref{eq:c2}), from this it follows
the invariance of the relations between the masses
under time transformation $t \rightarrow t+const$, $m_1/m_2=const$.
On the other hand, the scale of energy (frequency) evolves
in accordance with eq.~(\ref{eq:Et}), and the density of the matter
evolves in accordance with eqs.~(\ref{eq:q}), (\ref{eq:rhx}).
From this one observes the universe with the stationary
euclidean space and the stationary distributed matter.
The evolution of the universe manifests itself in the
change of the scale of energy (frequency)
or in the change of the density of the matter with time.

Let an observer view the universe at the modern time $t_0$ with
the modern size $a_0$.
Due to the evolution of the scale of energy~(\ref{eq:Et}),
the energy of photon decreases with time $E\propto 1/t$ and
correspondingly with the distance covered by photon $E\propto 1/r$.
In the model under consideration,
the radial and angular diameter distances are the same.
In view of eq.~(\ref{eq:Et}), the distance, radial or
angular diameter, is given by
\begin{equation}
\frac{r}{a_0}=\frac{\Delta E}{E}=\frac{z}{1+z}.
\label{eq:rz}
\end{equation}
Eq.~(\ref{eq:rz}) describes the Hubble law, with the
redshift being caused by the evolution of the scale
of energy.
In view of eq.~(\ref{eq:rhx}),
the intrinsic luminosity of the object $L$
and the observed flux $F$ are related as
\begin{equation}
F=\frac{L}{4\pi r^2(1+z)^2}=
\frac{L}{4\pi r_{\rm L}^2}
\label{eq:F}
\end{equation}
where
$r_{\rm L}$ is the luminosity distance.
The luminosity distance is expressed via
the angular diameter distance as
\begin{equation}
r_{\rm L}=r(1+z)=a_0 z.
\label{eq:rl}
\end{equation}

The change of the energy of photon with redshift can be registered
in two ways, as the change of the frequency of photon $\omega\propto 1+z$,
with the photon emission rate being fixed $1/t=const$,
or as the change of the photon emission rate $1/t\propto 1+z$,
with the frequency of photon being fixed $\omega=const$.
While measuring the photon flux through the photometric band,
one deals with the later case.
From this K-corrections which arise due to the change
of the frequency of photon with redshift have no meaning.

\subsection{Rescaling due to relativistic and quantum additions}

Eq.~(\ref{eq:rz}) holds for the local region $r\ll a_0$.
Taking into account the relativistic addition,
rewrite eq.~(\ref{eq:rz}) in the form
\begin{equation}
\frac{r}{a_0}=\frac{z}{1+z}-\frac{1}{2}\left(\frac{z}{1+z}\right)^2.
\label{eq:rz1}
\end{equation}
From eq.~(\ref{eq:rz1}) it follows that,
at $z\rightarrow 0$, $r \propto a_0$,
whereas, at $z\rightarrow\infty$, $r\rightarrow 1/2 a_0$.
Thus due to relativistic effects
the distance in the local region $z\rightarrow 0$
differs by a factor of 2
from the distance in the global region $z\rightarrow\infty$.
Introduce the local scale of time $\tau$ which is related to
the global scale of time as
\begin{equation}
\tau=\frac{1}{2}t.
\label{eq:tau}
\end{equation}
In view of eq.~(\ref{eq:mt}),
the local scale of time defines
the local scale of length and
the local scale of mass
\begin{equation}
a(\tau)=\frac{1}{2}a(t).
\label{eq:atau}
\end{equation}
\begin{equation}
m(\tau)=\frac{1}{2}m(t).
\label{eq:mtau}
\end{equation}

In the quantum electrodynamics, the scale of mass is normalized
with the electromagnetic coupling $\alpha$
\begin{equation}
m\propto \alpha.
\label{eq:mal}
\end{equation}
Due to quantum additions, the electromagnetic coupling $\alpha$
changes with momentum transferred. Since the energy scale in the universe
evolves it is necessary to take into account
the change of the scale of mass and equivalently of the scales
of length and time
due to the change of the electromagnetic coupling $\alpha$.

\section{Predictions and observational constraints}

\subsection{Constraints from the modern age and Hubble parameter}

Determine the modern age of the universe
from eq.~(\ref{eq:tem}) taking into account
the change of the electromagnetic coupling $\alpha$
due to quantum additions
\begin{equation}
t_{0}=\alpha_0 t_{Pl}\left(\frac{T_{Pl}}{T_{0}}\right)^2
\label{eq:age}
\end{equation}
where $T$ is
the temperature of the cosmic microwave background radiation,
the subscript $Pl$ corresponds to the Planck period,
the subscript $0$ corresponds to the modern period.
Here it is taken into account that $\alpha_{Pl}=1$.
Calculations yield the modern age of the universe
$t_0=33.7 \ {\rm Gyr}$. This value corresponds to the global scale.
In view of eq.~(\ref{eq:tau}),
the local value for the modern age of the universe is
$\tau_0=t_{0}/2=16.9 \ {\rm Gyr}$,
whereas the experimental value is
$\tau_{0}=14 \pm 2 \ {\rm Gyr}$~\cite{age}.
The modern age of the universe $\tau_0=16.9 \ {\rm Gyr}$ yields the
modern Hubble parameter $H_{0}=1/\tau_{0}=58{\ \rm km/s/Mpc}$,
whereas the experimental value is
$H_0=60\pm 10{\ \rm \ km/s/Mpc}$~\cite{Pat}.

\subsection{Constraints from the magnitude-redshift relation}

Two independent groups~\cite{Ia} have been published data on
the redshift-luminosity relation of SN~Ia at redshifts $z \sim 0.4 - 0.8$.
The difference in apparent magnitudes of
objects with the same intrinsic luminosity but at different redshifts
provides a valuable, classical cosmological test
on the evolution law of the universe.
The expected difference in apparent magnitudes is given by
$\Delta m \equiv 5log[r_{\rm L}(z_{2})/r_{\rm L}(z_{1})]$.
In view of eq.~(\ref{eq:rl}), for $z_{1} = 0.4$ and $z_{2} = 0.8$,
$\Delta m =1.5$.
As shown above, applying K-corrections have no meaning, so
it is necessary to use the magnitudes without K-corrections.
According to Perlmutter et al.~\cite{Ia},
without applying K-corrections
$\Delta m \approx 1.5 \pm 0.2$
that favours the model of the universe under consideration.
That is constraints from the magnitude-redshift relation of SN~Ia
favour the linear evolution law $a=t$.

\subsection{Constraints from primordial nucleosynthesis}

Authors of~\cite{Steig} investigated
constraints on power-law models of the universe,
in particular, from primordial nucleosynthesis.
The time-temperature relation $T^{-1}\leq t^{0.58}$
is obtained from the condition that at the beginning of
nucleosynthesis when $T\sim 80 {\ \rm keV}$ the age of the
universe should be less than
the lifetime of neutron $t\leq 887 {\ \rm s}$.
The inferred primordial abundances of helium-4~\cite{helium}
and deuterium~\cite{deut} require $T^{-1}\sim t^{0.55}$~\cite{Steig}.
Note that here $T$ is the temperature of the cosmic microwave
background radiation, $t$ is the age of the universe which are
related to their modern values.

Since the temperature of the cosmic microwave
background radiation is defined by the energy density,
according to eq.~(\ref{eq:tem}) the time-temperature relation
is $T^{-1} \sim t^{0.5}$.
This relation is obtained with the use of the global scales.
Constraints from primordial nucleosynthesis
limited by the lifetime of neutron $T^{-1}\leq t^{0.58}$ and
those limited by
the inferred primordial abundances of helium-4 and deuterium
$T^{-1} \sim t^{0.55}$ are
obtained with the use of the local scales.
It is necessary to take into account the transition factors
from the local scales to the global scales
for the modern age of the universe and
for the weak interaction rates in the primordial plasma which
define the lifetime of neutron and
the time to establish the neutron-proton equilibrium.
In view of eq. (\ref{eq:atau}),
the local-global transition factor
for the modern age of the universe is equal to 2.
The weak interaction rates
depend on energy as $\sim E^5$.
In view of eq.~(\ref{eq:mtau}),
the local-global transition factor
for the energy is equal to 2, and correspondingly
the local-global transition factor for
the weak interaction rates is equal to $2^5=32$.
Taking into account
the local-global transition factors
for the modern age of the universe and for
the weak interaction rates,
the power index in the relation
$T^{-1} \sim t^{0.5}$ increases by the value $\sim 0.05$.
Thus this time-temperature relation
is in agreement with the
constraints from primordial nucleosynthesis
limited by the lifetime of neutron $T^{-1}\leq t^{0.58}$ and
with those limited by
the inferred primordial abundances of helium-4 and deuterium
$T^{-1} \sim t^{0.55}$.

\subsection{The fractal galaxy distribution}

The galaxy distribution can be viewed as a fractal
(see e.g.~\cite{Pee}),
with the average number of galaxies within radius
$R$ from any given galaxy being given by
\begin{equation}
N(<R) \propto R^D
\label{eq:monop}
\end{equation}
where $D$ is the fractal power index.
The conventional point of view is that, on scales
$< 20\ h^{-1}\ {\rm Mpc}$, galaxies obey $D\approx 1.2-2.2$.
On scales $> 20\ h^{-1}\ {\rm Mpc}$, the fractal power index
increases with scale towards the value $D=3$ on scales of about
$100\ h^{-1}\ {\rm Mpc}$.
On the contrary authors of~\cite{Pi},\cite{slmp98}
claimed that galaxies have a fractal distribution
with constant $D\approx 2$ on all scales.
Having based on
the analysis of the galaxy number counts $N(<m)$,
they noted that
galaxy evolution,
modification of the Euclidean geometry and the K-corrections
are not very relevant in the range of the present data.
Authors of~\cite{Sc}
following~\cite{slmp98} reanalyzed the ESP survey
and suggested the value $D\simeq 3$.
But they reproduced the result of~\cite{slmp98}
for the case of neglecting K-corrections and using
euclidean metric with the luminosity distance $r_{\rm L}=a_0 z$.

In the universe under consideration,
the fractal galaxy distribution
arises due to the evolution of the scale of mass with time.
Write the average number of particles in the form
\begin{equation}
N(<R) \propto \rho R^3 /m
\label{eq:<R}
\end{equation}
where $m$ is the mass of the particle.
Define the fixed matter density $\rho =const$ at $t=const$
in all the regions of the universe. In view of eq.~(\ref{eq:mt}),
the scale of mass changes with time as $m\propto t$. Hence
the scale of mass changes with radius as $m\propto R$.
Taking into account the evolution of the scale of mass,
the average number of particles takes the form 
\begin{equation}
N(<R)\propto R^2.
\label{eq:<R1}
\end{equation}
Thus the universe under consideration predicts
the fractal galaxy distribution with the power index $D=2$.
Note that the case of neglecting K-corrections and using
euclidean metric with the luminosity distance $r_{\rm L}=a_0 z$
for which the ESP survey data show $D\approx 2$~\cite{Sc}
corresponds to the universe under consideration.

\end{document}